\documentstyle[preprint,aps]{revtex}
\begin{document}
\draft
\date{\today}     
\title{Was The Electromagnetic Spectrum A Blackbody Spectrum In The 
Early Universe?}
\author{Merav Opher\footnote{email: merav@orion.iagusp.usp.br}, 
Reuven Opher\footnote{email: opher@orion.iagusp.usp.br}}
\address{Instituto Astron\^omico e Geof\' \i sico - IAG/USP, Av. Miguel 
St\'efano, 4200 \\ 
CEP 04301-904 S\~ao Paulo, S.P., Brazil}

\maketitle

\begin{abstract}
It is assumed, in general, that the electromagnetic spectrum in the 
Primordial Universe was a blackbody spectrum in vacuum. We derive the 
electromagnetic spectrum, based on the {\it Fluctuation-Dissipation 
Theorem} that describes the electromagnetic 
fluctuations in a plasma. Our description includes thermal and collisional
effects in a plasma. The electromagnetic spectrum obtained differs from the 
blackbody spectrum in vacuum at low frequencies. In particular, concentrating 
on the primordial nucleosynthesis era, it has more energy for frequencies 
less than $3$ to $6\omega_{pe}$, where $\omega_{pe}$ is the electron plasma 
frequency.
\end{abstract}

\section{Introduction}
\label{sec:In}

It is usually assumed in Cosmology that the primordial plasma was 
a homogeneous plasma and that the electromagnetic field was a blackbody 
spectrum in vacuum. These assumptions are used, for example, in standard 
Big Bang Nucleosynthesis calculations. Deviations from a blackbody spectrum 
in vacuum and a homogeneous plasma can affect Primordial Nucleosynthesis and 
in this letter we concentrate on this epoch.

In the epoch of Big Bang Nucleosynthesis, the universe was a thermal 
bath of photons, electrons, positrons, baryons and neutrinos. The usual 
treatment considers the universe as a homogeneous plasma in thermal 
equilibrium and the electromagnetic field as a blackbody spectrum in vacuum. 
For example, in the energy density calculation, the energy density of the 
photons is given by the energy density of the blackbody spectrum in vacuum. 

In this manner, the Primordial Universe is treated as an ideal gas: {\it 
collective effects} are assumed to be negligible. 
However, a plasma differs from an ideal gas due to correlations, for example, 
two-particle correlations. In an ideal gas, the distribution function for 
two-particles is: $F_{2}(x_{1},x_{2})=F_{1}(x_{1})F_{2}(x_{2})$. In a plasma 
we have an additional term
$F_{2}(x_{1},x_{2})=[1+P_{12}(x_{1},x_{2})]F_{1}(x_{1})F_{2}(x_{2})$, where
$P_{12}$ is the Debye-Huckel screening. ($P_{12} 
\propto exp(-\mid x_{2} - x_{1} \mid/\lambda_{D})/\mid{x_{2}-x_{1}}\mid$, 
where $\lambda_{D}$ is the Debye length.)

A plasma, even in thermal equilibrium, has fluctuations, that is,
the physical variables such as temperature, density and electromagnetic fields,
fluctuate. Even for a non-magnetized plasma, where the average magnetic field 
is zero, $\langle B \rangle = 0$, the squared average is not zero: 
$\langle B^{2} \rangle \neq 0$. 

The study of the electromagnetic fluctuations in a plasma 
has been made in numerous studies, including those of Dawson, 
Rostoker et al., Sitenko and Gurin and Akhiezer et al. \cite{daw}. 
Most of the results are compiled in 
Sitenko and Akhiezer et al. \cite{sit}. 

The electromagnetic fluctuations are described by the {\it Fluctuation-
Dissipation Theorem} \cite{sit}. The intensity of such fluctuations is 
highly dependent on how the plasma is described, for example, 
on the dissipation mechanisms present in it. It is necessary to describe 
the plasma in the most complete way.

Cable and Tajima \cite{ct} (see also \cite{tc,t}) studied the magnetic 
field fluctuations, for several cases. Two of 
there descriptions concern the primordial plasma, which is an isotropic, 
non-magnetized and non-degenerate plasma. In particular, they studied: 
a) a cold, gaseous plasma and b) a warm, gaseous plasma described by 
kinetic theory.

In their study, Cable and Tajima \cite{ct} in case (a) used the 
{\it cold plasma} description with a constant collision frequency. In case 
(b) they analyzed the spectrum only for low frequencies, with the 
{\it warm plasma} description for phase velocity $\omega/k$ less or equal 
to the thermal velocity of the electrons, $v_{e}$, and the ions, $v_{i}$, in 
a collisionless description.

Through the study of the electromagnetic fluctuations, that is, the 
magnetic and electric field transverse fluctuations, given for example by 
the {\it Fluctuation-Dissipation Theorem}, a reliable 
electromagnetic spectrum can be obtained. 
In this letter, we present a model that extends the work of 
Cable and Tajima \cite{ct}. It includes in the same 
description collisional and thermal effects. By using the 
{\it Fluctuation-Dissipation Theorem} relations, we derive the 
electromagnetic spectrum in a plasma. We concentrated, in particular, in the primordial 
nucleosynthesis era, in the case of the electron-positron plasma at high temperatures and the 
electron-proton plasma at low temperatures. The first case is the plasma at the beginning of the 
primordial nucleosynthesis, when the number of electrons and positrons was comparable to the 
number of photons. The second case is the plasma at low temperatures, after the annihilation of 
the electrons and positrons where the number of electrons and protons was $\eta \sim 10^{-10}$ 
smaller than the number of photons.

In Section II we present the expression of the electromagnetic 
fluctuations, in Section III our model, and in Section IV a discussion 
of our results and conclusions.

\section{Electromagnetic Fluctuations}
\label{sec:mag}

The spectrae of fluctuations of the magnetic and electric transverse fields, 
for an isotropic plasma, given by the {\it Fluctuation-Dissipation Theorem} 
(for the deduction see Sitenko and Akhiezer et al. \cite{sit}), 
are 
\begin{equation}
\frac{{\langle B^{2} \rangle}_{{\bf k}{\omega}}}{8\pi}=2
\frac{\hbar}{e^{{\hbar}{\omega}/T}-1}
{\left( \frac{k c}{\omega} \right) }^{2}
\frac{Im~{\varepsilon}_{T}}
{{\mid {\varepsilon}_{T}-{\left( \frac{k c}{\omega}\right) }^{2} \mid}^2}~~;
~~
\frac{{\langle {E_{T}}^{2} \rangle}_{{\bf k}{\omega}}}{8\pi}=2
\frac{\hbar}{e^{{\hbar}{\omega}/T}-1}
\frac{Im~{\varepsilon}_{T}}
{{\mid {\varepsilon}_{T}-{\left( \frac{k c}{\omega}\right) }^{2} \mid}^2}~,
\label{em}
\end{equation}
where $\varepsilon_{T}$ is the transverse dielectric permittivity of the 
plasma.  

An intuitive way to understand the above expressions, is that the 
{\it Fluctuation-Dissipation Theorem} takes into account the emission 
and absorption processes in a plasma, and knowing that in 
equilibrium they are equal, the fluctuation 
level is obtained. 

To determine the fluctuations of the magnetic and electric transverse fields 
in a plasma in equilibrium or quasi-equilibrium, it is sufficient to know the 
transverse dielectric permittivity of the plasma, in particular, the 
dissipation mechanisms present for each frequency and wave number 
($Im~\varepsilon_{T}$). This depends on the treatment used to describe the 
plasma. 

Another important feature of the electromagnetic spectral distributions can 
be seen from Eq. (1). The equation 
$\varepsilon_{T}(\omega,{\bf k})-{\left ( \frac{kc}{\omega} \right)}^{2}=0$
determines the transverse eigenfrequencies of the plasma. 
In the transparency region ($Im~\varepsilon_{T} \ll Re~\varepsilon_{T}$), 
the magnetic and the electric field transverse spectrae have 
$\delta$-function-like maximae near the eigenfrequencies (i.e., 
the frequency spectrum of the fluctuations 
contains only the transverse eigenfrequencies in the plasma).
Knowing that photons have the dispersion relation in the plasma 
$\omega^{2}=\omega_{p}^{2}+k^{2}c^{2}$, we note that for frequencies 
$\omega \gg \omega_{p}$ the eigenfrequencies dominate
and the electromagnetic transverse field spectrum behaves like a
blackbody spectrum.

\section{BGK Collision Term}
\label{sec:mod}

To describe completely the plasma, we present a model that includes 
thermal effects as well as collisional effects. 
For this we need a kinetic description that takes into account 
collisions. We use the Vlasov equation in first order \cite{ich} 
(in the plasma parameter $g=1/n\lambda_{D}^{3}$, where $\lambda_{D}$ is the 
Debye length and $n$ is the particle density). The collisions are described 
by the collision term in this equation.

Charged particles simultaneously interact with all the particles 
in the Debye sphere, which is a large number and the Fokker-Planck collision 
term is most appropriate for a fully ionized plasma. It describes the effect of 
the microscopic fields produced by all the particles in the plasma.   
However, the kinetic equations with the Fokker-Planck collision term 
are very hard to solve. Therefore, we used the BGK collision term 
as a rough guide to the inclusion of collisions in the plasma.   
The BGK collision term describes the binary 
collisions as used by Cable and Tajima \cite{ct} and is given by,
(for a derivation see Clemnow and Dougherty or 
Alexandrov et al. \cite{cle}) 
\begin{equation}
\left ( \frac{\partial f}{\partial t} \right )_{C} = -\eta(f-f_{max})~,
\end{equation}
where $\eta$ is the collision frequency (considered constant and equal to
the Coulomb collision frequency). $f_{max}$ is
given by $f_{max}({\bf x},{\bf v},t)=N({\bf x},t)f_{0}({\bf v})/{N_{0}}$, 
where the number density is $N=N_{0}+N_{1}$ and $f=f_{0}+f_{1}$, $f_{0}$ 
being the unperturbed Maxwellian distribution. (This collision term conserves 
the number of particles.) 
Substituting this collision term in the equation of Vlasov in first order, it 
is straightforward to obtain, for an isotropic plasma, the transverse dielectric 
permittivity $\varepsilon_{T}$, (generalized for several species):
\begin{equation}
\varepsilon_{T}(\omega,{\bf k}) = 1 + \sum_{\alpha} 
\frac{{\omega_{p\alpha}}^{2}}{\omega^{2}}
\left ( \frac{\omega}{\sqrt{2}kv_{\alpha}} \right ) Z \left ( 
\frac{\omega+i\eta_{\alpha}}{\sqrt{2}kv_{\alpha}} \right )~,
\label{etw}
\end{equation}
where $\alpha$ is the label for each species of the plasma, $v_{\alpha}$
the thermal velocity for each species and $Z(z)$ the Fried $\&$ Conte
function \cite{fc}, 
\begin{equation}
Z(z)=\frac{1}{\sqrt{\pi}}\int_{-\infty}^{\infty} \frac{dt~e^{t^2}}{t-z}~.
\end{equation}
(If relativistic temperature effects are included, the substitution 
$\omega_{p\alpha} \rightarrow \omega_{p\alpha}/\sqrt{\gamma}$ is made.)

It is interesting to comment and emphasize that both the {\it cold plasma} 
description and the {\it warm plasma} collisionless description, used 
by Cable and Tajima \cite{ct}, are particular solutions of this model.  For 
${\mid z \mid}^{2} \gg 1$, where $z=({\omega+i\eta})/{\sqrt{2}kv_{e}}$, we 
obtain the {\it cold plasma} dielectric permittivity and for 
$\eta \rightarrow 0$ the {\it warm plasma} collisionless dielectric 
permittivity.

We substitute the transverse dielectric permittivity in Eq. (1), 
obtaining the magnetic and the electric transverse field spectrum, 
${\langle B^{2} \rangle}_{k\omega}$ and 
${\langle {E_{T}}^{2} \rangle}_{k\omega}$, respectively. Integrating over 
wave number (and dividing by  $(2\pi)^{3}$) gives us the magnetic and 
electric field transverse spectrae, 
${\langle B^{2} \rangle}_{\omega}$ and 
${\langle {E_{T}}^{2} \rangle}_{\omega}$, respectively.

Our model uses a kinetic theory description with a collision term that describes the binary 
collisions in the plasma and a cut-off has to be taken, since for very small distances the energy of the Coulomb 
interactions of the particles exceeds their kinetic energy which violates the 
applicability of the condition of the perturbation expansion (in the plasma 
parameter $g \ll 1$). This occurs approximately for distances 
$r_{min} \sim  e^{2}/T$, or more exactly, the distance of closest approach 
between a test particle and an electron 
in a plasma, $r_{min}={(k_{max})}^{-1}\cong {[{Mmv^{2}}/{(m+M)} \mid eq\mid ]}^{-1}$, where 
$M$, $v$ and $q$ are respectively, the mass, velocity and charge of the test 
particle \cite{ich}. 

Treating properly the effects of distant encounters, 
the cut-off procedure can be removed. This was shown by    
Thompson and Hubbard, and Hubbard, in several works [9], analyzing the 
Fokker-Planck equation and its coefficients. 
They showed that the cut-off procedure is unnecessary, when higher order terms 
in the Fokker-Planck equations are maintained. The kinetic equations with the
Fokker-Planck collision term are very hard to solve, however. We used
the BGK collision term where a cut-off is necessary. We took $k_{max}$
consistent with this collision term. 

The electromagnetic spectrum is obtained by 
summing the magnetic and electric field transverse spectrae:
\begin{equation}
S({\omega})=
\frac{{\langle B^{2} \rangle}_{\omega}}{8\pi} + 
\frac{{\langle {E_{T}}^{2} \rangle}_{\omega}}{8\pi}~.
\label{spec}
\end{equation}

In Figure 1a, we plot the electromagnetic spectrum 
$S({\omega})$ given by Eq. \ref{spec} (divided by the 
normalization $S_{0}=\omega_{pe}^{2}k_{B}T/c^{3}$) vs $\omega/\omega_{pe}$. 
We study an electron-positron plasma at $T=10^{10}~K$ and 
$n_{e}=1.4\times10^{31}~cm^{-3}$. 
This is the plasma at the beginning of the Primordial Nucleosynthesis era, 
when neutrinos decoupled from the plasma and the neutron to proton ratio become 
frozen-in. (This ratio essentially determines the primordial $^{4}He$
abundance.) The dotted curve is $S(\omega)$ (our model) compared to the blackbody 
spectrum in vacuum (the solid curve). In Figure 1b, we plot the same curves 
as in Figure 1a, but extended to high frequencies. In Figure 1c and 1d, we 
did the same as Figure 1a and 1b, but for an electron-proton plasma with 
$T=10^{9}~K$ and $n_{e}=5.4\times 10^{26}$. In the epoch of Primordial 
Nucleosynthesis, at lower temperatures, the electrons and positrons 
annihilate and the plasma is reduced to a plasma of protons and electrons.

It can be seen that we obtain the blackbody spectrum naturally for high 
frequencies, and in the case of the high-temperature plasma ($T=10^{10}~K$), 
for frequencies $\omega \leq 3~\omega_{pe}$, the spectrum 
has more energy than the blackbody spectrum. In the case of the low-
temperature plasma ($T=10^{9}~K$) it has more energy for frequencies 
$\omega \leq 6\omega_{pe}$.

\section{Discussion and Conclusions}
\label{sec:con}

The unique manner to obtain the electromagnetic transverse spectrum is 
analyzing the magnetic and electric field transverse fluctuations. This is 
the only manner to obtain information, not only about modes that propagate, 
like photons, but also modes that do not propagate. These modes appear, not 
only at low frequencies but also at high frequencies, resulting 
from the correlations in the plasma. Only at very high frequencies, the 
photons contribute uniquely to the magnetic and electric field transverse 
spectrae. 

We present a model that incorporates, in the same description, the thermal 
and collisional effects and used the {\it Fluctuation-Dissipation 
Theorem} that describes the electromagnetic fluctuations. We use the Vlasov 
equation with the BGK collision term. 

The final electromagnetic spectrum for the primordial plasma at the epoch 
of Big Bang Nucleosynthesis behaves like a blackbody spectrum in vacuum for 
high frequencies. However, for low frequencies, it is distorted. It has 
more energy than the blackbody spectrum in vacuum. In the case of the 
high-temperature plasma ($T=10^{10}~K$), this range is for frequencies 
$\omega \leq 3\omega_{pe}$ and for the low-temperature plasma ($T=10^{9}~K$) 
the range is for $\omega \leq 6\omega_{pe}$, where $\omega_{pe}$ is the 
electron plasma frequency. 

This additional energy is due to the collective modes of the plasma. The reason 
why the collective modes of the plasma can have more energy for 
$\omega \leq \omega_{pe}$ than the photons in vacuum, can be 
understood as follows. Photons are massless bosons with the dispersion 
relation $\omega^{2}=k^{2}c^{2}$. For the energy interval 
$0 \leq \omega \leq \omega_{pe}$, the wave number interval is $k=0$ to 
$k=\omega_{pe}/c$. A relatively small amount of phase space is involved. 
For the collective motions of the plasma, in general, we have a larger amount 
of phase space. For example, for plasmons with energy 
$\omega \sim \omega_{pe}$, the amount of phase space extends to a maximum $k$ 
of $k_{D} \cong \omega_{pe}/v_{T}$, where $v_{T}$ is the thermal electron 
velocity, which is greater than $\omega_{pe}/c$ for the 
photons. In general, for a given frequency for $\omega < \omega_{pe}$, the 
greater phase space available to the collective 
modes of the plasma (than that of the photons) implies more energy, or a higher 
spectrum. 

This result (the additional energy that appears in the 
electromagnetic spectrum compared to the blackbody energy usually assumed) 
can affect several fields in Cosmology, in particular, Big Bang 
Nucleosynthesis. This extra energy, that has not been previously taken into 
account, causes the Universe to expand more rapidly at a given temperature. 
In particular, it causes the neutrinos to decouple earlier (at a higher 
temperature) and the neutron to proton ratio to freeze-in at a higher 
value. (In order to estimate the total additional energy, we have to add also, the longitudinal 
energy due to the longitudinal electric field spectrum. For $T=0.8~MeV$, we 
obtain for the additional energy,  
$\Delta \rho \cong 1\% \rho_{\gamma}$, where $\rho_{\gamma}=\rho_{BB}$, the energy density of the 
blackbody photon spectrum in vacuum. A complete study, estimating the additional energy for 
diverse temperatures and densities, is in preparation \cite{op}).

Another interesting aspect is how the additional energy affects the spectrum of the microwave 
background. The 
additional energy at $z_{DEC}$ (the redshift when the spectrum was formed) occurred at 
frequencies $\sim 10^{-9}~\omega_{peak}$ ($\omega_{peak}=2.8k_{B}T/\hbar$). However, non-linear effects in 
plasma can bring the additional energy to higher frequencies. This is very interesting and should be 
investigated in the future.

The authors would like to thank Swadesh Mahajan for useful suggestions, 
especially concerning the BGK collision term. The authors also would like to 
thank Arthur Elfimov for helpful discussions and the anonymous referees for
helpful comments. M.O. would like to thank the
Brazilian agency FAPESP for support and R.O. would like to thank the 
Brazilian agency CNPq for partial support.

\newpage
\begin{figure}
\caption{The electromagnetic spectrum 
$ln[S(\omega)/S_{0}]$ vs $\omega/\omega_{pe}$
(where $S(\omega)={\langle B^{2} \rangle}_{\omega}/8\pi + 
{\langle {E_{T}}^{2} \rangle}_{\omega}/8\pi$ and 
$S_{0}={\omega_{pe}}^{2}k_{B}T/c^{3}$ is the normalization) for: 
(a) The electron-positron plasma at $T=10^{10}~K$ and 
$n_{e}=1.4\times 10^{31} cm^{-3}$
(the dotted curve is {\it our model} and 
the solid curve is the blackbody spectrum in vacuum); 
(b) The same as case (a), extended to high frequencies;  
(c) The same as case (a) but for an electron-proton plasma at 
$T=10^{9} K$ and $n_{e}=5.4\times 10^{26}$; and
(d) The same as case (c), extended to high frequencies.} 
\label{fig1}
\end{figure}


\begin{thebibliography}{60}

\bibitem{daw} 
J. M. Dawson, Adv. Plasma Phys., {\bf 1}, 1 (1968); 
N. Rostoker, R. Aamodt, and O. Eldridge, Ann. Phys. {\bf 31}, 243 (1965); 
A. G. Sitenko, and A. A. Gurin, JETP {\bf 22}, 1089 (1966); 
A. I. Akhiezer, I. A. Akhiezer, and Sitenko, A. G, JETP {\bf 14}, 462 (1961). 

\bibitem{sit}
A. G. Sitenko, {\it Electromagnetic Fluctuations in Plasma}
(Academic Press, NY, 1967); 
A. I. Akhiezer, I. A. Akhiezer, R. V. Plovin, A. G. Sitenko, 
and K. N. Stepanov, {\it Plasma Electrodynamics}, Vol. 2 (Pergamon Press, 
Oxford, 1975).

\bibitem{ct} 
S. Cable, and T. Tajima, \pra {\bf 46}, 3413 (1992). 

\bibitem{tc} 
T. Tajima, S. Cable, and R. M. Kulsrud, Phys. Fluids B {\bf 4}, 2338 (1992).

\bibitem{t} 
T. Tajima, S. Cable, K. Shibata, and R. M. Kulsrud, \apj {\bf 390}, 309 
(1992).

\bibitem{cle} 
P. C. Clemmow, and J. P. Dougherty,  
{\it Electrodynamics of Particles and Plasmas}, 
(Addison-Wesley, Redwood City, CA, 1990); 
A. F. Alexandrov,  L. S. Bogdankevich, and  A. A. Rukhadze, 
{\it Principles of Plasma Electrodynamics} 
(Springer-Verlag, Berlin, 1984).

\bibitem{fc} 
B. D. Fried, and S. D. Conte, {\it The Plasma Dispersion Function} 
(Academic Press, NY, 1961).

\bibitem{ich} 
S. Ichimaru, {\it Basic Principles of Plasma Physics}  
(Addison-Wesley, Redwood City, CA, 1992).

\bibitem{th}
W. B. Thompson and J. Hubbard, Rev. Mod. Phys. {\bf 32}, 714 (1960); 
J. Hubbard, Proc. Roy. Soc. A {\bf 260}, 114 (1961); 
J. Hubbard, Proc. Roy. Soc. A {\bf 261}, 371 (1961).

\bibitem{op}
M. Opher and R. Opher (in preparation).

\end{thebibliography}
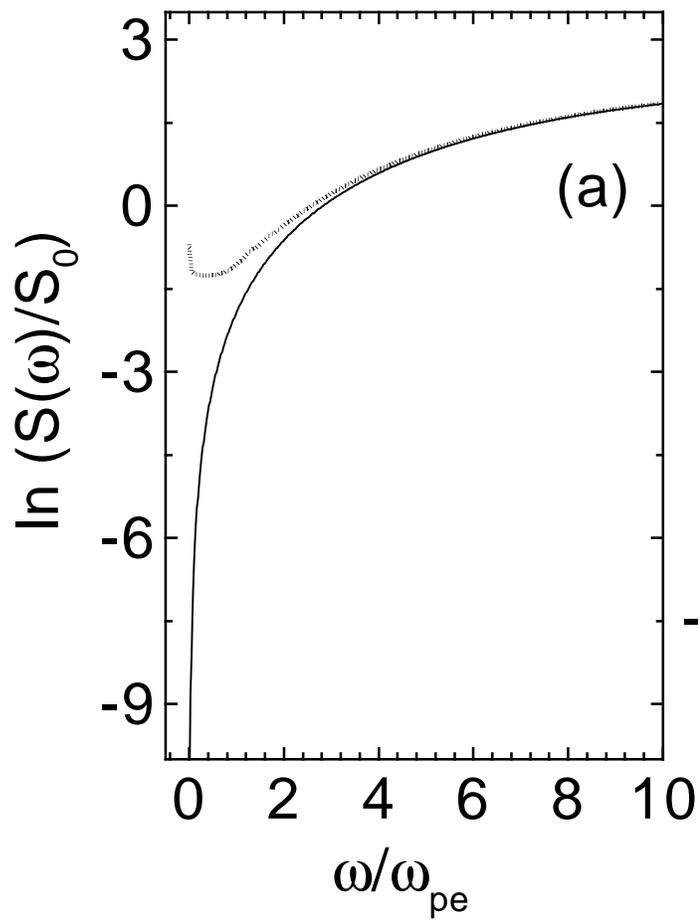
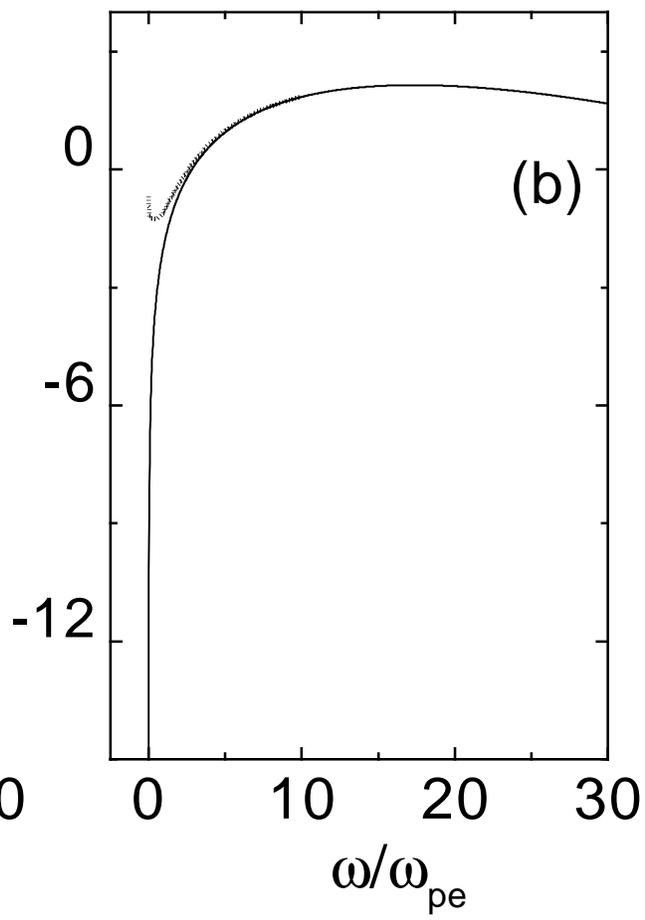
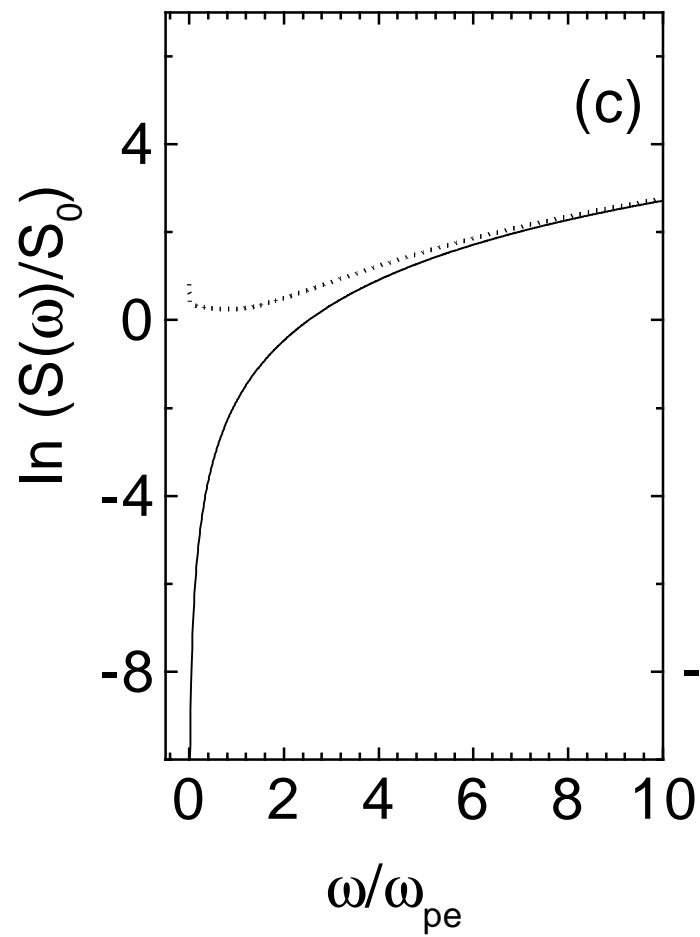
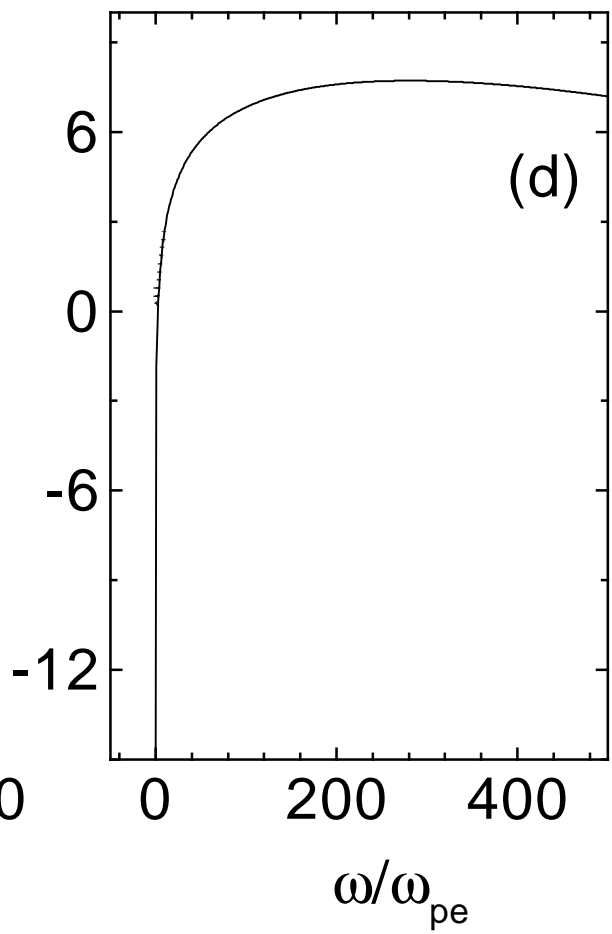
\end{document}